\documentstyle[aps]{revtex}
\begin{document}
\title{Realistic Shell-Model Calculations for Nuclei 
in the Region of Shell Closures off Stability}
\author{A. Covello, L. Coraggio, and A. Gargano}
\address{Dipartimento di Scienze Fisiche, Universit\`a 
di Napoli Federico II, \\ and Istituto Nazionale di Fisica Nucleare, \\
Complesso Universitario di Monte S. Angelo, Via Cintia - 80126 Napoli, Italy}

\date{\today}
\maketitle
\begin{abstract}

We have performed realistic shell-model calculations
for nuclei around doubly magic $^{100}$Sn and $^{132}$Sn using an
effective interaction derived from the Bonn A 
nucleon-nucleon potential. The results are in remarkably good
agreement with the experimental data showing the ability of
our effective interaction to provide an accurate description
of nuclear structure properties.   

\end{abstract}

\draft

\section{Introduction}

A fundamental goal in nuclear structure theory is to understand the
properties of complex nuclei in terms of the nucleon-nucleon ($NN$) interaction.
Since the pioneering work of Kuo and Brown 
\cite{KB}, who in the mid 1960s derived an {\it s-d} shell effective 
interaction from the Hamada-Johnston potential \cite{HJ}, 
there has been considerable progress in this field. On the one hand, the
theoretical framework in which the model-space effective interaction
$V_{\rm eff}$ can be derived from a given ($NN$) potential has
been largely improved (the main aspects of this derivation are reviewed
in Ref. \cite{Kuo96}). On the other hand, high-quality $NN$ potentials
have been constructed which give a very good description of the $NN$
scattering data. A comprehensive review of modern $NN$ potentials
suitable for application in nuclear structure is given in 
Ref. \cite{Machl94}. 

These improvements have brought about a great deal of renewed interest
in realistic shell-model calculations.
In this context, the main question is to which extent modern realistic
interactions can provide a consistent and accurate description of
nuclear structure phenomena. To try to answer this question, the study
of nuclei in the vicinity of closed shell is extremely important.
In fact, they provide the best testing ground for the basic ingredients
of a shell-model calculation, in particular as regards the matrix
elements of the effective $NN$ interaction.

In this paper, we shall present some results of realistic shell-model
calculations for nuclei around doubly magic $^{100}$Sn and $^{132}$Sn.
They are the $N=50$ isotones $^{98}$Cd, $^{97}$Ag, $^{96}$Pd, the $N=82$
isotones $^{134}$Te and $^{135}$I, and the doubly odd nucleus $^{132}$Sb,
which has a single proton outside the $Z=50$ closed shell 
and a single neutron hole in the closed $N=82$ shell.
These nuclei lie well away from the valley of stability and only
recently more experimental information has become available which
is of great value for a stringent test of our calculated effective
interaction. It should be noted that the study of $^{132}$Sb provides
a direct test of the effective interaction matrix elements with
isospin $T=0$.

In earlier works we have performed shell-model calculations for the
light Sn isotopes making use of two effective interactions
derived from different free $NN$ potentials. They are the
Paris  \cite{Lacom80} and Bonn A \cite{Machl87} potentials. 
It turned out that the latter, which has a weaker tensor force component,
leads to the best agreement with experiment for all of the nuclei
considered \cite{Andr96,Cov97}. For this reason, we have used it
in the present work. 

Our presentation is organized as follows. In Sect. 2 we give a brief
description of our calculations. In Sect. 3 we present our results and
compare them with the experimental data. Sect. 4 presents a summary
of our conclusions.

\section{Outline of calculations}
As already mentioned in the Introduction, we make use of a two-body
effective interaction derived from the meson theoretic Bonn A potential. 
This was obtained using a $G$-matrix folded-diagram
formalism, including renormalizations from both core polarization and
folded diagrams. A description of the derivation of the effective
interaction $V_{\rm eff}$ from the nucleon-nucleon potential including
references can be found in Ref. \cite{Kuo96}. 
We only outline here the essential of the method and point out the
main differences between our present and earlier calculations
\cite{Andr96,Cov97,Andr97}. 

The effective interaction can be schematically written \cite{Kuo90}
in operator form as
$$V_{\rm eff} = \hat{Q} - \hat{Q}' \int \hat{Q} + \hat{Q}' \int \hat{Q} \int
\hat{Q} - \hat{Q}' \int \hat{Q} \int \hat{Q} \int \hat{Q}\, \cdots \; ,
\eqno (1) $$
where $\hat{Q}$ and $\hat{Q'}$ represent the $\hat{Q}$-box, composed of
irreducible valence-linked diagrams, and the integral sign represents a
generalized folding operation. We take the $\hat{Q}$-box to be
composed of $G$-matrix diagrams through second-order in $G$;
they are just the seven first- and second-order diagrams
considered by Shurpin {\it et al.} \cite{Shurp83}.
To calculate the effective interaction $V_{\rm eff}$ given by Eq. (1),
a first step is to calculate the irreducible $\hat Q$-box diagrams and
their energy derivatives in terms of the model-space $G$-matrix defined
by \cite{Kren76}
$$G(\omega)=V+V Q_2 {{1} \over
{\omega-Q_2TQ_2}} Q_2G(\omega) . \eqno (2)$$
Here  $Q_2$ is the Pauli exclusion operator for the two interacting
nucleons, V represents the $NN$ potential, T denotes the two-nucleon kinetic 
energy, and $\omega$ is the so-called starting energy.
We employ a matrix inversion method to calculate the above $G$ matrix
in an essentially exact way \cite{Kren76}. 

For the $N=50$ isotones, we have considered the doubly closed $^{100}$Sn
as an inert core and treated protons as valence holes. This leads to
a calculation of the $\hat{Q}$-diagrams which is different from the usual
one for particles. A detailed description of the calculation of our
two-hole effective interaction will be given in a forthcoming
publication. We have chosen the Pauli exclusion operator
$Q_2$ in the $G$-matrix equation (2) as specified \cite{Kren76} by 
($n_1, n_2, n_3$) = (11, 21, 45). For the shell-model oscillator 
parameter $\hbar \omega$
we have used the value 8.5 MeV, as obtained from the expression
$\hbar \omega = 45 A^{-1/3} - 25A^{-2/3}$ for $A=100$.

For the $N=82$ isotones the valence-proton and -neutron
orbits outside the $^{132}$Sn core are different.  
Therefore, we have chosen ($n_1, n_2, n_3$) = (16, 28, 55)  for the
neutron orbits and ($n_1, n_2, n_3$) = (11, 21, 55) for the proton orbits.
The effective interaction has been calculated using
an isospin uncoupled representation, where neutrons and protons are
treated separately. The adopted value of oscillator spacing is 
$\hbar \omega=7.88$ MeV.  
It should be mentioned here that in our earlier study \cite{Andr97} we
made the choice ($n_1, n_2, n_3$) = (11, 28, 45) for both protons and
neutrons. As a consequence, the Pauli exclusion operator was not treated
in a completely correct way.

As regards $^{132}$Sb, we have treated the odd proton and the
remaining 31 neutrons as valence particles. This makes 
the $T=1$ matrix elements of the effective interaction be just 
the same as those used in our earlier study of the Sn isotopes \cite{Andr96},
to which we refer the reader for details.

For the $N=82$ isotones and for $^{132}$Sb our model space includes
the five single-particle (s.p.) orbits $0g_{7/2}$, $1d_{5/2}$, $2s_{1/2}$, 
$1d_{3/2}$, and $0h_{11/2}$, while for the $N=50$ isotones the proton
holes are distributed in $0g_{9/2}$, $1p_{1/2}$, $1p_{3/2}$, 
and $0_{5/2}$ orbits.

As regards the choice of the s.p. energies, we have proceeded as follows.
For the $N=82$ isotones we have taken three s.p. spacings from the
experimental spectrum of $^{133}$Sb \cite{Serg86,Sanchez98}. In fact, 
the $g_{7/2}$, $d_{5/2}$, $d_{3/2}$ and 
$h_{11/2}$ states can be associated with the ground state and the
0.962, 2.439 and 2.793 MeV excited levels, respectively. 
It should be noted that the position of the $d_{3/2}$ state
corresponds to the new value very recently provided by the
high-sensitivity $\gamma$ spectroscopic measurement of Ref.
\cite{Sanchez98}.
This is about 400 keV smaller than the previously accepted value
(2.708 MeV) \cite{Serg86}. It turns out, however, that this 
change has little effect on the spectra of $^{134}$Te and $^{135}$I.
As for the
$s_{1/2}$ state, its position has been determined by reproducing the
experimental energy of the ${1 \over 2}^+$ level at 2.150 MeV in
$^{137}$Cs. This yields the value $\epsilon_{s_{1/2}}$ = 2.8 MeV.

For the $N=50$ isotones, the single-hole energies cannot be taken
from experiment, since the single-hole valence nucleus $^{99}$In 
has not yet been studied. Therefore we have determined them from
an analysis of the low-energy spectra of the isotones with $A \geq 92$.
The adopted values (in MeV) are:
$\epsilon_{g_{9/2}}$ = 0.0, $\epsilon_{p_{1/2}}$ = 0.73, $\epsilon_{p_{3/2}}$
= 2.17, $\epsilon_{f_{5/2}}$ = 3.24.

Regarding $^{132}$Sb, we have assumed the s.p. energies to be the same
for neutrons and protons.
Our adopted values (in MeV) are: $\epsilon_{d_{5/2}}=0$,
$\epsilon_{g_{7/2}}=0.20$, $\epsilon_{s_{1/2}}=1.72$, $\epsilon_{d_{3/2}}=1.88$,
and $\epsilon_{h_{11/2}}=2.70$. As compared to the set of s.p. energies used 
for the Sn isotopes (see Ref. \cite{Andr96}), only $\epsilon_{s_{1/2}}$ 
and $\epsilon_{d_{3/2}}$ have been
modified. More precisely, they have been both decreased by about 0.5 MeV.
It should be noted that the position of these two
levels plays a minor role in the calculations for the light Sn isotopes
while it is very important to satisfactorily reproduce the 
experimental $1_2^+$ state and to place in the right energy range 
the calculated negative-parity states.

\section{Results}

\subsection {$N=50$ isotones}

In Fig. 1 we compare the calculated $2^+$, $4^+$, $6^+$ and $8^+$
yrast states of $^{98}$Cd with those recently identified in the study 
of Ref. \cite{Gorska97}.
We see that our results are in very good agreement with
experiment. 
A measure of the quality of the agreement between theory 
and experiment is given by the rms deviation $\sigma$ \cite{sigma},
whose value is 105 keV.

It should be noted that the predicted position of
the $5^-$ state (2.73 MeV) is quite consistent with the experimental 
information available for the two lighter even isotones. In fact,
in $^{96}$Pd and $^{94}$Ru a $5^-$ state has been observed 
\cite{Peker93,Tuli92} at 2.648 and 2.625 MeV, respectively. 
As regards the structure of
the excited states, we find that the positive-parity states are
of practically pure ($\pi g_{9/2}$)$^{-2}$ character while the
$5^-$ state is dominated by the  $\pi g_{9/2}^{-1}p_{1/2}^{-1}$
configuration. On the contrary, the ground-state has
a significantly mixed wave function, the percentage of configurations
other than ($\pi g_{9/2}$)$^{-2}$ being about 17\%.

In Figs. 2 and 3 the calculated spectra of $^{97}$Ag and $^{96}$Pd 
are compared with the experimental ones \cite{Artna93,Peker93}.
We see that for both nuclei our calculations produce a spectrum very 
close to the experimental one. 
The main points of disagreement are the position of the 
${{13} \over 2}^+$ state in $^{97}$Ag and that of the
$2^+$ state in $^{96}$Pd which lie 213 and 281 keV, respectively, 
above the observed ones. The $\sigma$ value is 121 keV and 130 keV
for $^{97}$Ag and $^{96}$Pd, respectively. 

Experimental information on electromagnetic transition rates in heavy
$N=50$ isotones is very scanty. It is of interest to mention here,
however, that in the work of Ref. \cite{Gorska97} a value
of 0.44$^{+0.20}_{-0.10}$ W.u. for the $B(E2; 8^+ \rightarrow 6^+)$  
in $^{98}$Cd has been reported. Using the bare proton charge
$e_p=1$ we obtain 0.67 W.u. By contrast, to reproduce the experimental
value of the same $B(E2)$ in $^{96}$Pd ($0.34 \pm 0.05$ W.u.)
an effective charge
of at least $1.7e$ is needed. To clear up this point more experimental
data are required. 

\subsection{$^{134}$Te and $^{135}$I}

The experimental \cite{Omt95,Serg94} and theoretical spectra of the two-proton
nucleus $^{134}$Te are compared in Fig. 4, where all the calculated 
and experimental levels up to 3.2 MeV excitation energy are reported. 
We see that while the theory reproduces all the observed levels it also 
predicts the existence of a $3^+$ and a $0^+$ state at 2.65 and 2.78 MeV, 
respectively. This prediction is strongly supported by the experimental 
information available for the two heavier even isotones. In fact, 
in $^{136}$Xe a $0^+$ state has been observed \cite {Tuli94} at 2.58 MeV 
while in $^{138}$Ba both a $0^+$ and a $3^+$ state have been located
\cite{Tuli93} at 2.34 and 2.45 MeV , respectively. 
Above 3.2 MeV excitation energy the comparison between theory 
and experiment is made only for those observed levels which have received a 
spin-parity assignment. We do not include the new 
levels observed in \cite{Zhang96} since all of them 
should be interpreted as neutron particle-hole states. This interpretation 
is confirmed by our calculations. 
In fact, the only two states, having $J^\pi= 8^+$ and $10^+$, which can 
be constructed in our model space are both predicted to lie at about 7.3 
MeV while the two experimental states with these spin-parity assignments have 
been located \cite{Zhang96} at 4.557 and 5.622 MeV, respectively. 
We see that the calculated spectrum reproduces
very well the experimental one, the discrepancies between theory and
experiment being smaller than 50 keV for several states (9 out of 15). The rms 
deviation $\sigma$ is 106 keV. 

In Fig. 5 we compare the calculated spectrum of $^{135}$I with the experimental
one \cite{Zhang96,Serg87} up to 4.0 MeV excitation energy. As in the case 
of $^{134}$Te, we exclude the experimental levels above 4.2 MeV, which 
originate from core excitations. We should note that the spectra of 
Fig. 5 include all experimental and calculated levels up to 1.5 MeV.
Above this energy several other levels without 
assigned spin and parity are reported in \cite{Serg87}; we compare our calculated 
states only with those observed in \cite {Zhang96}. From Fig. 5 we see that the 
excitation energies are remarkably well reproduced for all the reported states,
the $\sigma$ value being 58 keV. We have associated the experimental level at 
1.010 MeV with the theoretical ${3\over2}^+$ at 0.931 MeV. 

We should now point out that the theoretical results presented here for  
$^{134}$Te and $^{135}$I are in a substantially better agreement with
experiment than those obtained in our earlier study \cite{Andr97}.
The reason for these improvements can be clearly traced to 
the better treatment of the Pauli exclusion operator $Q_2$.

In Table I we compare the experimental reduced transition probabilities
in $^{134}$Te with the calculated ones. We have used an effective proton
charge $e_p^{\rm eff}=1.55 e$. This is consistent with the values adopted by other
authors \cite{Wilden71,Heyde82}. The theoretical $B(E2)$ values are in very good
agreement with experiment. As regards the $E3$ transitions, we find that
the $B(E3;9_1^- \rightarrow 6^+_2)$ is well reproduced while the
$B(E3;9_1^- \rightarrow 6^+_1)$ is underestimated by a factor of about
4. A possible reason for this discrepancy lies in the fact that only
a small amount of configuration mixing is present in the calculated
$6^+$ states. In fact, the decay to the $6_2^+$ state is dominated
by the single-proton transition  $(h_{11/2}g_{7/2})_{9^-} \rightarrow
(g_{7/2}d_{5/2})_{6^+}$ while that to the $6^+_1$ state by the transition
$(h_{11/2}g_{7/2})_{9^-} \rightarrow (g^2_{7/2})_{6^+}$, which
is retarded owing to spin flip. The theoretical 
$B(E3;9_1^- \rightarrow 6^+_1)$ value could be brought in agreement with
experiment by an amount of configuration mixing of about 15\%, 
which would, of course,
reduce the $B(E3;9_1^- \rightarrow 6^+_2)$ value. It should be noted,
however, that the latter would still be within the error bar.

\subsection{$^{132}$Sb}

The experimental and theoretical spectra of $^{132}$Sb are compared 
in Fig. 6, where all the
observed levels are reported. In the calculated spectrum all levels 
up to 1.4 MeV excitation energy are included while in the higher energy
region only the $1^+_2$ and $3^-_2$ states are reported. It should be 
noted that the
nature of the presently available experimental information is quite
different for positive- and negative-parity levels. In fact, while the
spin-parity assignments to the former have been clarified by the study of Ref.
\cite{Mach95}, this is not the case for the latter. More precisely,
the excitation energy of the $8^-$ state is not known (the work of
Ref. \cite{Stone89} places it between 150 and 250 keV) and the three other
observed negative-parity states have not received firm spin assignments.
We find that the first excited $8^-$ state lies at 126 keV while the
$6^-_1$, $5^-_1$, $7^-_1$, $3^-_1$, and $4^-_1$ states are grouped 
in a very small energy interval (from 210 to 380 keV). As a consequence,
any attempt to establish a one-to-one correspondence between the
observed levels and those predicted by our calculation could be
misleading. It is of interest to note that the
above states, which all arise from the $\pi g_{7/2} \nu h_{11/2}$
configuration, are well separated from the
other two members of the multiplet, i.e. the $9^-$ and $2^-$
states, which are predicted at 1.0 and 1.42 MeV,
respectively. A similar behavior is also predicted for the
$\pi d_{5/2} \nu h^{-1}_{11 /2}$ multiplet. In fact, the  
$7^-$, $6^-$, $5^-$, and $4^-$ states belonging to
this configuration lie between 0.82 and 0.97 Mev while the highest- and
lowest-spin members ($8^-$ and $3^-$) are at 1.12 and 1.56 MeV,
respectively.

From Fig. 6 we see that the experimental excitation energies of the
positive-parity states are remarkably well reproduced by the theory,
the largest discrepancy being 77 keV for the $5^+_1$ state. The value of the
rms deviation $\sigma$ is only 32 keV. 

In Table II we compare the $BE(2)$ values for transitions between
states below 1.1 MeV excitation energy with the calculated ones. 
A more complete analysis of the electromagnetic properties of
$^{132}$Sb may be found in \cite{Andr98}. We have used
an effective proton charge $e_p^{\rm eff}=1.55e$, which is
just the same as that adopted for the $N=82$ isotones. No effective charge
has been attributed to the neutron hole.
As we see from Table II, the experimental data are
affected by large errors. In view of this, the
agreement between theory and experiment can be
considered quite satisfactory. In fact, our calculated values lie all
but two within the limits set by experiment. 

\section{Concluding remarks}

We have presented here some recent results of realistic shell-model
calculations for nuclei around doubly magic $^{100}$Sn and $^{132}$Sn.
They have been obtained by employing an effective interaction derived
from the Bonn A nucleon-nucleon potential. We have shown that the
agreement between theory and experiment is very good for all
nuclei considered. It is to be emphasized that the 
study of $^{132}$Sb provides a test of our $T=0$ effective 
interaction in this mass region. This is of special interest since in 
earlier works using different $NN$ potentials it turned out that not 
enough attraction was provided by the calculated matrix elements of the 
$T=0$ effective interaction, which has a stronger dependence on the 
tensor force strength than the $T=1$ interaction (a detailed discussion 
of this important point including references is given in Ref. 
\cite{Jiang92}).

In conclusion, the success achieved by our calculations shows
that our effective interaction derived from
the Bonn A $NN$ potential is able to describe
with quantitative accuracy the spectroscopic properties of 
nuclei near closed shells.

%\newpage
\acknowledgments
The results presented in this paper are part of a research program
carried out in collaboration with F. Andreozzi, T. T. S. Kuo, and
A. Porrino.
This work was supported in part by the Italian Ministero dell'Universit\`a
e della Ricerca Scientifica e Tecnologica (MURST).

\begin{table}
\caption{Calculated and experimental $B(E\lambda)$ values (in W.u.) 
for $^{134}$Te. The experimental data are from [20].}
\begin{tabular}{llld}
$J^\pi_i \rightarrow J^\pi_f$ & $\lambda$ & $B(E\lambda)_{\rm expt}$ 
& $B(E\lambda)_{\rm calc}$ \\
\tableline
$4^+_1 \rightarrow 2^+_1$ & 2 & ${4.3 \pm 0.3}$ & 4.2  \\
$6^+_1 \rightarrow 4^+_1$ & 2 & ${2.05 \pm 0.03}$ & 1.9  \\
$9^-_1 \rightarrow 6^+_1$ & 3 & ${3.8 \pm 0.2}$ & 1.0  \\
$9^-_1 \rightarrow 6^+_2$ & 3 & ${8.0 \pm 1.3}$ & 8.2  \\
\end{tabular}
\end{table}

\begin{table}
\caption {Calculated and experimental $B(E2)$ values (in $e^2$ fm$^4$) 
for $^{132}$Sb. The experimental data are from [28].}
\begin{tabular}{lcd}
$J^\pi_i \rightarrow J^\pi_f$ &  $B(E2)_{\rm expt}$ 
& $B(E2)_{\rm calc}$ \\
\tableline
$3^+_1 \rightarrow 4^+_1$ &  36$ \pm 11$ & 42  \\
$2^+_1 \rightarrow 3^+_1$ &  76$ \pm 76$ & 53  \\
$2^-_1 \rightarrow 4^+_1$ &  ${< 26}$      & 5.8  \\
$2^+_2 \rightarrow 3^+_2$ &  ${1.1^{+15.4}_{-1.1}}$ & 0.85 \\
$2^+_2 \rightarrow 2^+_1$ &  37$ \pm 30$  & 4.4  \\
$2^+_2 \rightarrow 3^+_1$ &  36$ \pm 23$  & 5.4  \\
$2^+_2 \rightarrow~4^+_1$ &  8.4$ \pm 4.5$  & 9.1  \\
\end{tabular}
\end{table}

\begin{figure}
\caption{Experimental and calculated spectrum of $^{98}$Cd.}
\label{fig. 1}
\end{figure}
\begin{figure}
\caption{Experimental and calculated spectrum of $^{97}$Ag.}
\label{fig. 2}
\end{figure}
\begin{figure}
\caption{Experimental and calculated spectrum of $^{96}$Pd.}
\label{fig. 3}
\end{figure}
\begin{figure}
\caption{Experimental and calculated spectrum of $^{134}$Te.}
\label{fig. 4}
\end{figure}
\begin{figure}
\caption{Experimental and calculated spectrum of $^{135}$I.}
\label{fig. 5}
\end{figure}
\begin{figure}
\caption{Experimental and calculated spectrum of $^{132}$Sb.}
\label{fig. 6}
\end{figure}


\begin{references}
\bibitem{sigma} We define $\sigma = \{ { {1 \over {N_d} }
\sum_i {[E_{\rm exp}(i) - E_{\rm calc}(i)]}^2 } \}^{1/2}  $, where $N_d$ is the
number of data.
\end{references}
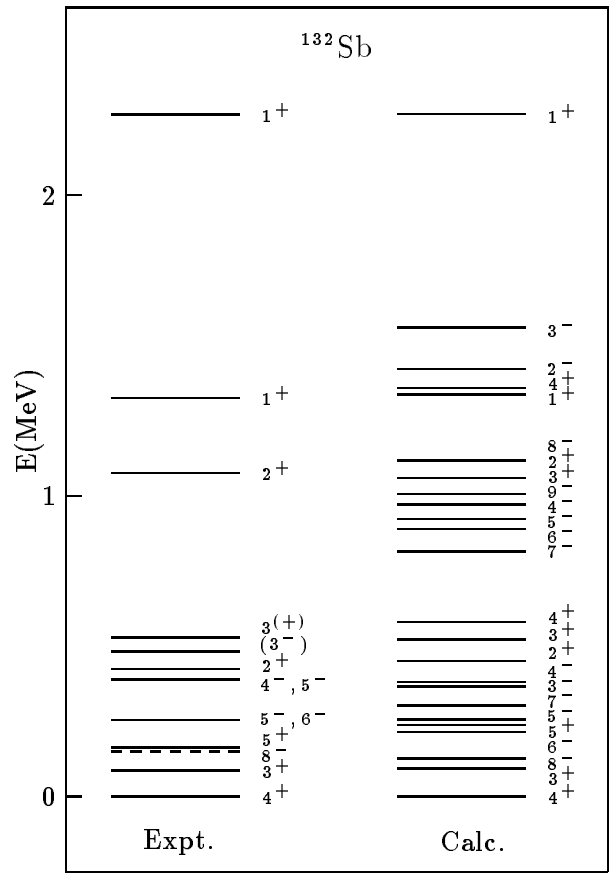
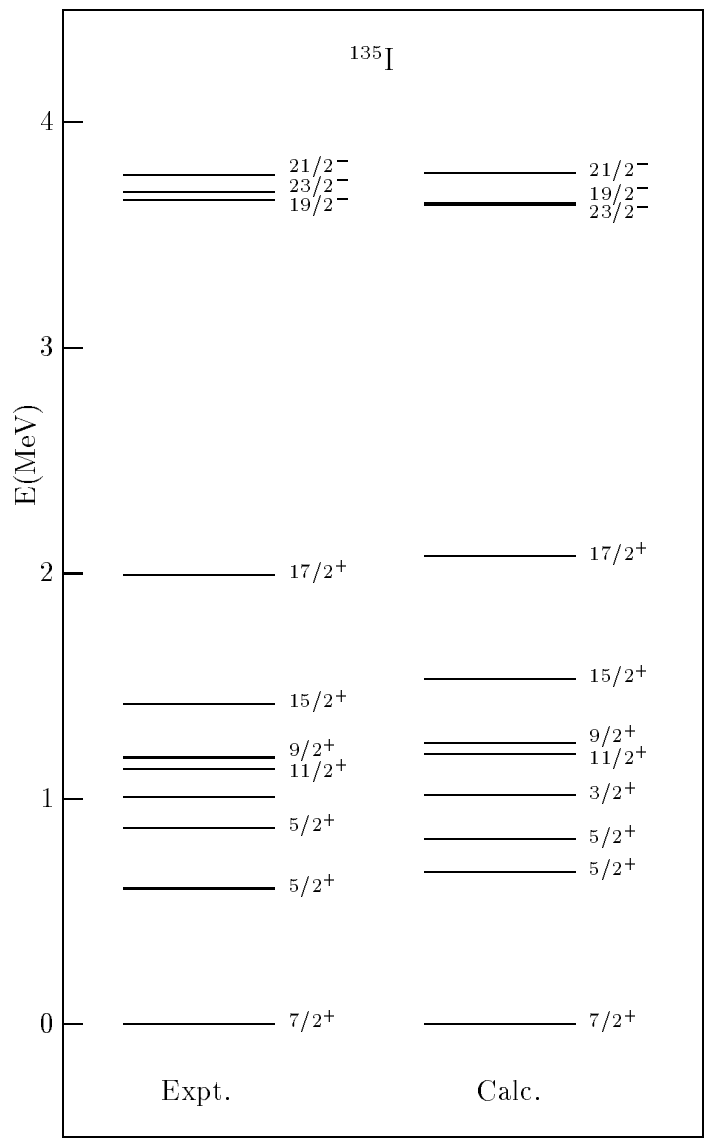
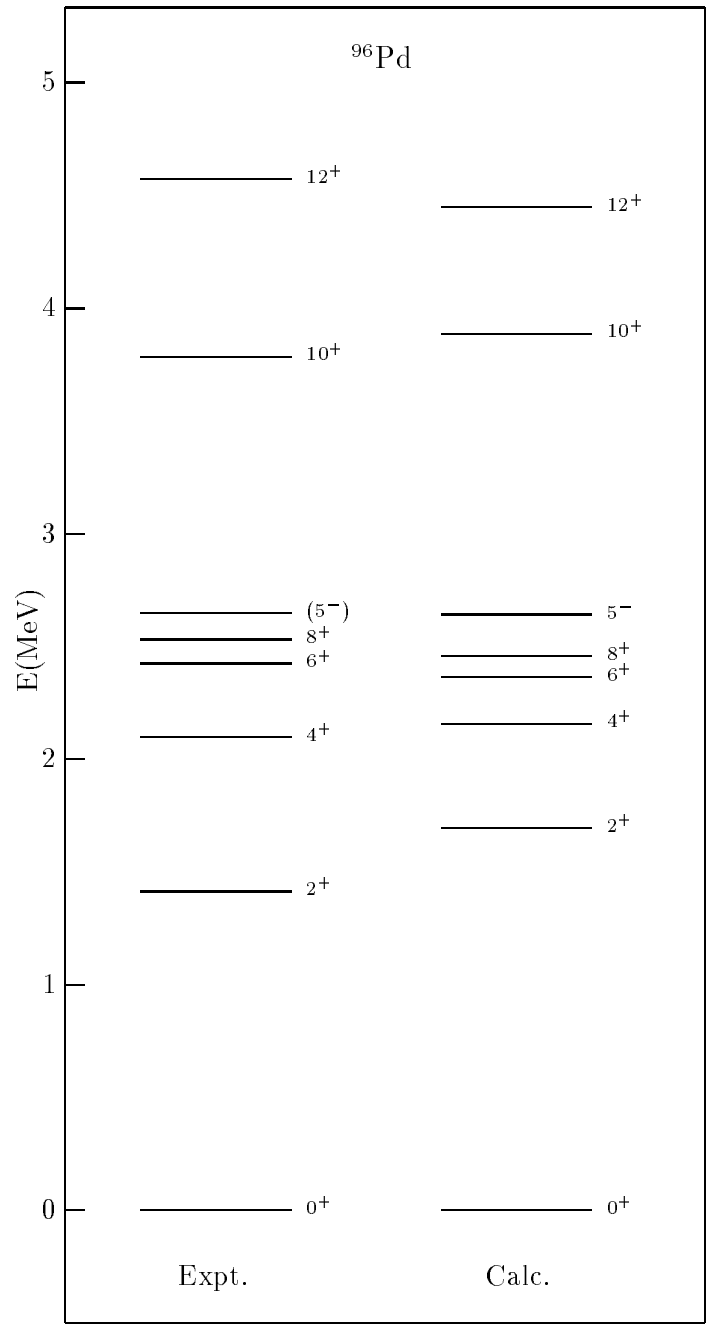
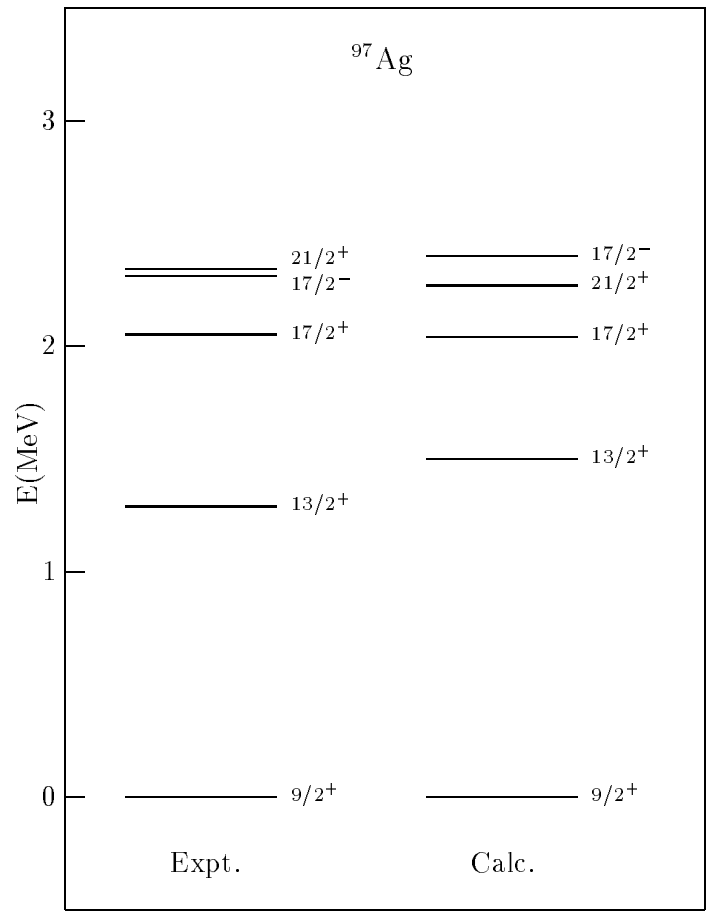
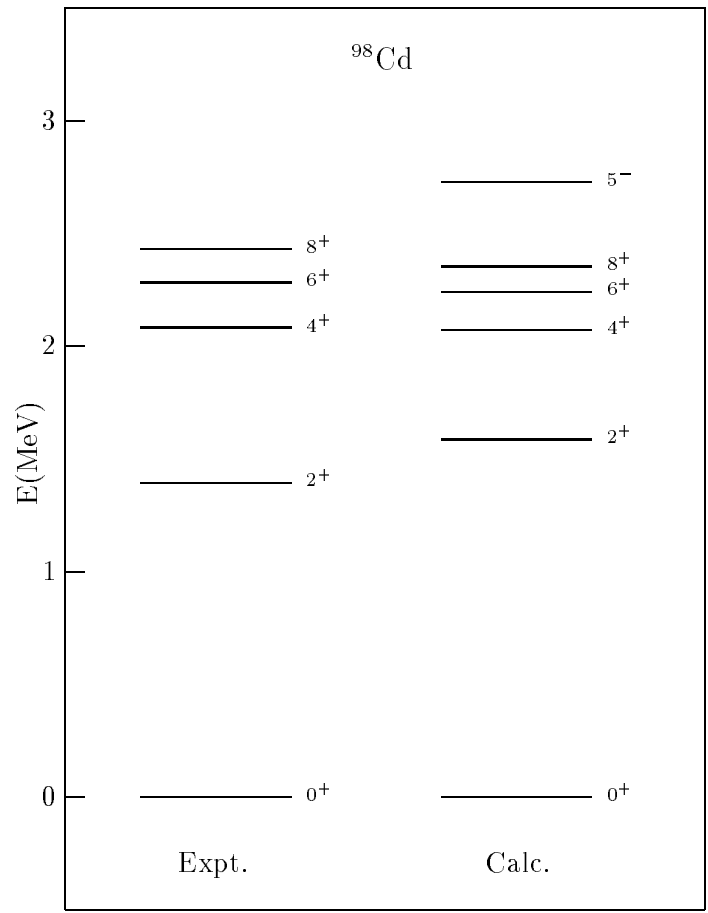
\end{document}